# High-Performance Depletion/Enhancement-Mode $\beta$-Ga$_2$O$_3$ on Insulator (GOOI) Field-Effect Transistors With Record Drain Currents of 600/450 mA/mm

Hong Zhou, Mengwei Si, Sami Alghamdi, Gang Qiu, Lingming Yang, and Peide D. Ye, *Fellow, IEEE*

*Abstract*—In this letter, we report on high-performance depletion/enhancement-mode $\beta$-Ga$_2$O$_3$ on insulator (GOOI) field-effect transistors (FETs) with record high drain currents ($I_D$) of 600/450 mA/mm, which are nearly one order of magnitude higher than any other reported $I_D$ values. The threshold voltage ($V_T$) can be modulated by varying the thickness of the $\beta$-Ga$_2$O$_3$ films and the E-mode GOOI FET can be simply achieved by shrinking the $\beta$-Ga$_2$O$_3$ film thickness. Benefiting from the good interface between $\beta$-Ga$_2$O$_3$ and SiO$_2$ and wide bandgap of $\beta$-Ga$_2$O$_3$, a negligible transfer characteristic hysteresis, high $I_D$ ON/OFF ratio of $10^{10}$, and low subthreshold swing of 140 mV/decade for a 300-nm-thick SiO$_2$ are observed. E-mode GOOI FET with source to drain spacing of 0.9-$\mu$m demonstrates a breakdown voltage of 185 V and an average electric field (E) of 2 MV/cm, showing the great promise of GOOI FET for future power devices.

*Index Terms*— $\beta$-Ga$_2$O$_3$, GOOI FET, D-mode, E-mode, nano-membrane.

## I. Introduction

MONOCLINIC $\beta$-Ga$_2$O$_3$ with an ultra-wide bandgap of 4.6-4.9 eV has been identified as a promising contender for the next generation power devices [1]–[8]. Its ultra-wide bandgap enables the $\beta$-Ga$_2$O$_3$ material to possess a critical breakdown field ($E_c$) of 8 MV/cm. Even at such early development stage, high average E of 3.8 MV/cm and high breakdown voltage (BV) of 750 V have already been achieved [2], [9]. Combined with 100 cm$^2$/V·s electron mobility ($\mu$) at room temperature, $\beta$-Ga$_2$O$_3$ possesses a high Baliga's figure of merit of 3444, defined as $\varepsilon\mu E_c^3$, where $\varepsilon$ is the dielectric constant of $\beta$-Ga$_2$O$_3$ [10]. In addition to its excellent material property, potential cost effective substrate can be realized through Czochralski method [11], [12]. Besides those aforementioned material characteristics, $\beta$-Ga$_2$O$_3$ crystal also possesses some unique properties. For instance, its (100) surface has a large lattice constant of 12.23 Å along [100] direction, which allows

Manuscript received November 7, 2016; revised November 26, 2016; accepted November 29, 2016. Date of publication December 2, 2016; date of current version December 27, 2016. This work was supported by the AFOSR under Grant FA9550-12-1-0180 and in part by DTRA under Grant HDTRA1-12-1-0025. The review of this letter was arranged by Editor M. Radosavljevic.
The authors are with the Birck Nanotechnology Center, School of Electrical and Computer Engineering, Purdue University, West Lafayette, IN 47907 USA (e-mail: yep@purdue.edu).
Color versions of one or more of the figures in this letter are available online at http://ieeexplore.ieee.org.
Digital Object Identifier 10.1109/LED.2016.2635579

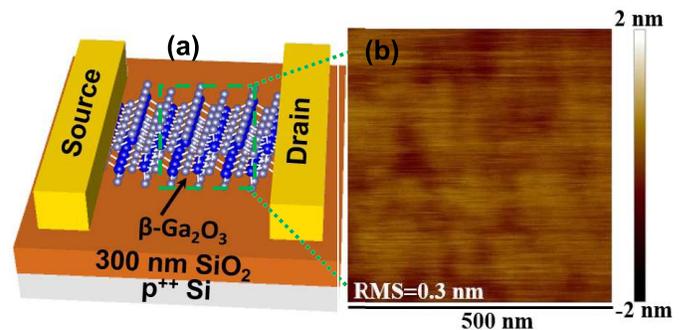

Fig. 1. (a) Schematic view of a GOOI FET with a 300 nm SiO$_2$ layer on Si substrate and (b) AFM image of the atomic flat $\beta$-Ga$_2$O$_3$ surface after cleavage.

a facile cleavage into thin belts or nano-membranes [6], [13]. Therefore, $\beta$-Ga$_2$O$_3$ on insulator (GOOI) field-effect transistor (FET) can be formed by transferring the $\beta$-Ga$_2$O$_3$ nano-membrane to SiO$_2$/Si substrate and followed by regular device fabrication.

On the other hand, the quest for the low on-resistance ($R_{on}$) and high $I_D$ are always demanded for improved power device performance. This situation is more severe in E-mode or normally-off devices, and currently the $I_D$ is less than 10 mA/mm with large $R_{on}$ [14], [15]. In this letter, we have successfully demonstrated high performance D-mode and E-mode GOOI FETs with record high $I_D$, record low $R_{on}$, high on/off ratio, low subthreshold swing (SS), and negligible hysteresis. Specifically, the record high performance E-mode GOOI FET, which satisfies the failure-safe requirement of power devices, can be simply achieved by optimizing the $\beta$-Ga$_2$O$_3$ thickness upon specs.

## II. Device Fabrication and Measurement

Fig. 1(a) and (b) are the schematic of a GOOI FET and atomic force microscopy (AFM) image of the $\beta$-Ga$_2$O$_3$ surface after cleavage, which shows atomically flat and uniform within the whole nano-membrane or the single device. Device fabrication was started from a 6 mm by 6 mm (-201) $\beta$-Ga$_2$O$_3$ bulk substrate with Sn doping concentration of $2.7 \times 10^{18}$ cm$^{-3}$, determined by capacitance-voltage (C-V) measurements [16]. Thin $\beta$-Ga$_2$O$_3$ nano-membrane was transferred from the substrate cleavage to the SiO$_2$/p$^+$ Si substrate with SiO$_2$ thickness of 300 nm. The SiO$_2$/Si substrates were cleaned in acetone for 24 hours and the Ga$_2$O$_3$ nano-membrane transfer time was within 1 minute. Then source and drain regions were defined by the VB6 e-beam lithography followed by Ti/Al/Au (15/60/50 nm) metallization and lift-off processes. Prior to the metal deposition, Ar plasma bombardment for 30 s was applied to generate oxygen vacancies to enhance the surface n-type doping for the reduction of the contact resistance ($R_c$). The device without Ar bombardment shows poor contacts with Schottky contact like behaviors. Meanwhile, the increase of Ar bombardment time over the optimized 30 s in our experiments results in an increase of the $R_c$. Various $\beta$-$Ga_2O_3$ nano-membranes with thickness from 50 nm to 150 nm, confirmed by the AFM measurements, were chosen for the device fabrication. During the device fabrication, there was no capping or protection layer on the $\beta$-$Ga_2O_3$ nano-membranes. The device characterizations were carried out with Keithley 4200 Semiconductor Parameter Analyzer.

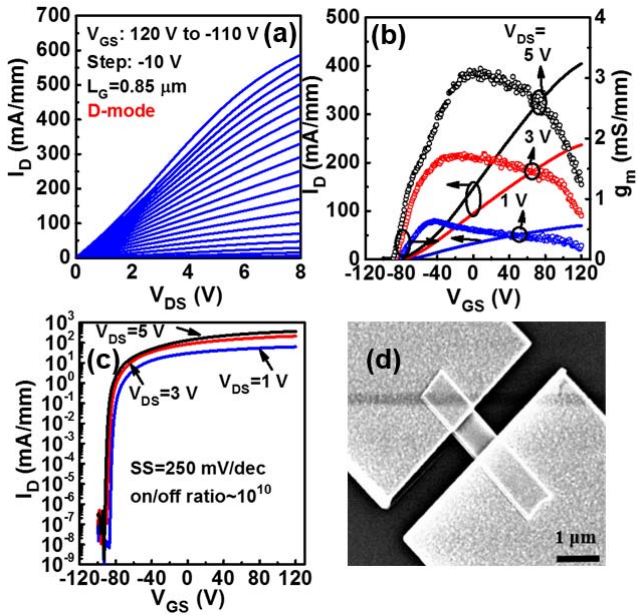

Fig. 2. (a) $I_D$-$V_{DS}$ output characteristics of a D-mode GOOI FET with 94 nm thick $\beta$-$Ga_2O_3$ nano-membrane. Record high maximum $I_D$ of 600 mA/mm is demonstrated. (b) and (c) are linear-scale $I_D$-$g_m$-$V_{GS}$ and log-scale $I_D$-$V_{GS}$ transfer characteristics of the same device with $V_T = -80$ V, respectively. High on/off ratio of $10^{10}$ and low SS = 250 mV/dec for 300nm $SiO_2$ are obtained. (d) SEM image of a fabricated D-mode GOOI FET.

The success of integrate $\beta$-$Ga_2O_3$ on Si substrate shows the potential to migrate the issue of low thermal conductivity of $\beta$-$Ga_2O_3$ substrate by wafer bonding $\beta$-$Ga_2O_3$ on AlN or diamond substrates. The advantage of this device fabrication process can enable to study $\beta$-$Ga_2O_3$ channel thickness dependent $V_T$, and provide a higher bandgap material underneath the $\beta$-$Ga_2O_3$ channel for BV enhancement. Most importantly, it offers an effective route to study the fundamental transport properties of $\beta$-$Ga_2O_3$ and device performance potentials without using many expensive $\beta$-$Ga_2O_3$ epitaxy wafers with different channel thickness.

## III. RESULTS AND DISCUSSION

Fig. 2(a) shows the well-behaved direct current (DC) output current-voltage (I-V) of a D-mode GOOI FET with source to drain spacing ($L_{SD}$, also gate length $L_G$) of 0.85 $\mu$m and channel thickness of 94 nm. The typical range of physical width of these nano-membrane devices is 0.3~1 $\mu$m, accurately determined by scanning electron microscopy (SEM) as shown in Fig. 2(d). Considering on the depletion width at the edges of the nano-membrane, the presented drain current densities are under-estimated because they are normalized by the physical width instead of effective electrical width. The measurements start from applying the back-gate bias ($V_{GS}$) to 120 V and then stepping to the device pinch-off $-110$ V with $-10$ V as the step, while the drain bias ($V_{DS}$) is swept from 0 to 8 V. Maximum drain current densities ($I_{DMAX}$) of 600 mA/mm is obtained, which is nearly one order of magnitude higher than any other reported $\beta$-$Ga_2O_3$ MOSFETs [1]–[9], [13]. We ascribe this record high $I_{DMAX}$ to the much higher doping concentration of $\beta$-$Ga_2O_3$ membrane applied and the positive back-gate bias reduced source and drain $R_c$. The $R_{on}$ is extracted to be 13 $\Omega$·mm. The $R_C$ and sheet resistance ($R_{SH}$) of D-mode devices are extracted to be 2.7 $\Omega$·mm and 8.5 k$\Omega$/□ at $V_{GS} = 120$ V, respectively, through the transfer length method (TLM) of various similar $\beta$-$Ga_2O_3$ thickness but with different $L_G$. The Schottky-like contacts with large $R_c$ lead to the $I_D$-$V_{DS}$ output characteristics of the D-mode devices showing curvature in the linear region. More efforts are needed to improve the contacts, i.e. by Si or Sn ion implantation, to further boost the device performance. Fig. 2(b) and (c) are linear and log-scale transfer characteristics ($I_D$-$V_{GS}$) of the same device. A $V_T$ of $-80$ V is extracted from the linear extrapolation of $I_D$-$V_{GS}$ at $V_{DS} = 1$ V. The peak transconductance ($g_{max}$) is calculated to be 3.3 mS/mm at $V_{DS} = 5$ V. The peak field-effect mobility ($\mu_{FE}$) is calculated to be 48.8 $cm^2$/V·s from the $g_{max}$, which is still a factor of 2 lower than the theoretical limit [17]. Although the oxide thickness is 300 nm, high on/off ratio of $10^{10}$ and low SS of 250 mV/dec are obtained, showing the interface between $\beta$-$Ga_2O_3$ and $SiO_2$ is of high quality.

Fig. 3(a) presents the $I_D$-$V_{DS}$ output characteristics of an E-mode GOOI FET with $L_G = 1.3$ $\mu$m and channel thickness of 79 nm. A record high $I_{DMAX} = 450$ mA/mm is obtained, which is more than one order of magnitude higher than any other E-mode MOSFETs [14], [15]. Similar to D-mode device, the $R_{on}$, $R_{SH}$ and $R_C$ of E-mode device are extracted to be 20 $\Omega$·mm, 14.1 k$\Omega$/□, and 0.95 $\Omega$·mm, respectively. The reduced $R_c$ of E-mode devices is likely due to the thinner nano-membrane. The backage bias can be more effectively to electrostatically dope the thinner nano-membrane surface where the metal contacts are physically contacted. Fig. 3(b) and 3(c) are the linear and log-scale $I_D$-$g_m$-$V_{GS}$ plots of the same E-mode device. $V_T$ and $g_{max}$ of 7 V and 4.5 mS/mm are extracted at $V_{DS} = 1$ and 9 V, respectively. The peak $\mu_{FE}$ of E-mode device is calculated to be 55.2 $cm^2$/V·s from the $g_{max}$. High on/off ratio of $10^{10}$ and low SS = 140 mV/dec are also obtained for the E-mode device, benefiting from the ultra-wide bandgap of $\beta$-$Ga_2O_3$ and high quality interface. The much smaller SS of E-mode devices is due to the thinner $\beta$-$Ga_2O_3$. Fig. 3(d) depicts the $I_D$-$V_{GS}$ hysteresis measurements when the $V_{GS}$ is first swept from $-15$ V to 100 V and then swept back of another E-mode device with a similar thickness of 80 nm. There is a negligible hysteresis for the dual sweep $I_D$-$V_{GS}$ transfer curves, which further confirms the high quality interface between the $\beta$-$Ga_2O_3$ and $SiO_2$.

To have a direct comparison about the $V_T$ shift from negative values in D-mode to positive values in E-mode by reducing $\beta$-$Ga_2O_3$ nano-membrane thickness, we have carried out measurements on GOOI FETs with various $\beta$-$Ga_2O_3$ thickness. Fig. 4(a) describes the thickness dependent representative $I_D$-$V_{GS}$ characteristics. Obviously, the $V_T$ is shifted from negative to positive when the thickness is slowly reduced. Fig. 4(b) summarizes the extracted thickness dependent $V_T$ of 15 devices. Generally, they all follow the same trend as shown in Fig. 4(a). The determined thickness dependent $V_T$ may be valuable in the realization of high performance top gate E-mode GOOI FETs in the near future [18].

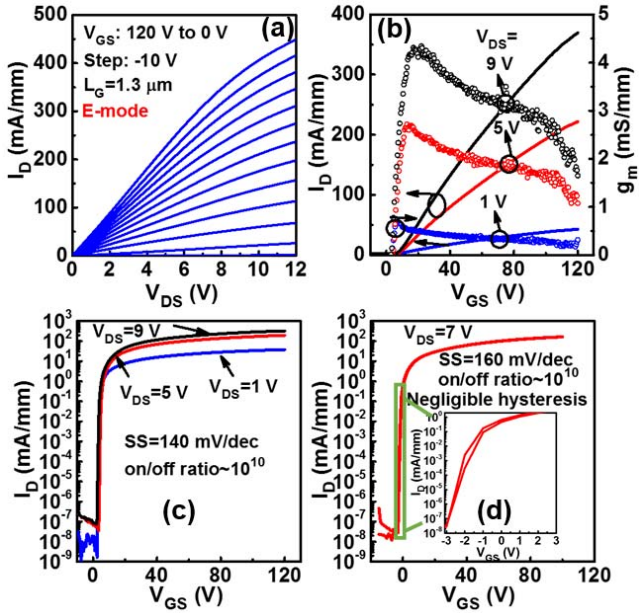

Fig. 3. (a) Output characteristics $I_D$-$V_{DS}$ of an E-mode GOOI FET with 79 nm thick of $\beta$-Ga$_2$O$_3$ nano-membrane. Record high maximum $I_D$ of 450 mA/mm is demonstrated. (b) and (c) Linear-scale $I_D$-$g_m$-$V_{GS}$ and log-scale $I_D$-$V_{GS}$ transfer characteristics of the same device with $V_T = 7$ V, respectively. High on/off ratio of $10^{10}$ and low SS = 140 mV/dec are obtained. (d) Dual-sweep hysteresis measurement of another device with thickness of 80 nm. Negligible hysteresis is observed, which shows the high quality interface.

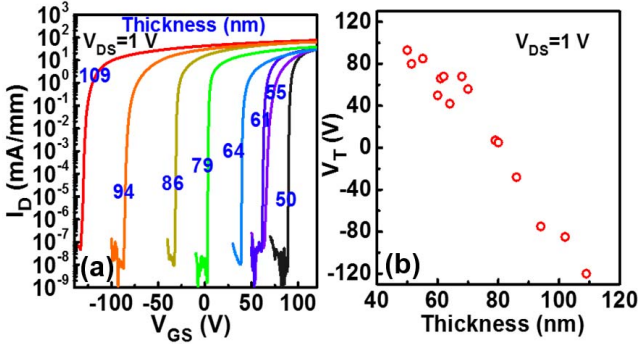

Fig. 4. (a) Thickness dependent $I_D$-$V_{GS}$ plots of various GOOI FETs from D-mode of thicker $\beta$-Ga$_2$O$_3$ to E-mode of thinner $\beta$-Ga$_2$O$_3$. (b) Thickness dependent $V_T$ extracted at $V_{DS} = 1$ V of 15 devices.

To evaluate the potential of GOOI FETs for power device applications, we have performed off-state breakdown measurements on E-mode device. Fig. 5(a) presents the off-state breakdown measurement of an E-mode GOOI FET with $L_{SD} = 0.9$ $\mu$m and the membrane thickness of 61 nm. The p$^+$ Si back-gate is floated during the measurement. The origin of the abrupt increase in $I_{off}$ at $V_{DS} = 60$ V is unclear. It is not representative since some of the measured devices don't have this change. A BV = 185 V is observed for the short $L_{SD} = 0.9$ $\mu$m. Compared with the BV = 230 V of the device from AFRL with $L_{SD}/L_{GD} = 4.4/0.6$ $\mu$m reported in [9], our work has reached nearly a 5 times lower $R_{on}$, which can potentially improve the thermal management issues during the on-state. There is a tradeoff between the BV and the thickness of SiO$_2$. Thick SiO$_2$ can help to increase the BV but it makes the poor thermal conductivity issue of $\beta$-Ga$_2$O$_3$ even worse. In the near future, GOOI FETs on AlN or diamond substrate might be a good solution to have high BV while maintaining high thermal conductivity of $\beta$-Ga$_2$O$_3$ to substrate. Fig. 5(b) shows the simulation of the electrostatic potential of the same device as Fig. 5(a). The $\beta$-Ga$_2$O$_3$ channel is modeled with n-type doping concentration of $1 \times 10^{13}$ cm$^{-3}$ to simulate the situation of $10^{-3} \sim 10^{-4}$ mA/mm off-state $I_D$. The simulated potential against position is plotted in Fig. 5(c). The average electrical field ($E_{av}$) in the channel is calculated to be 2 MV/cm, which further confirms the potential of GOOI FETs as next generation power devices.

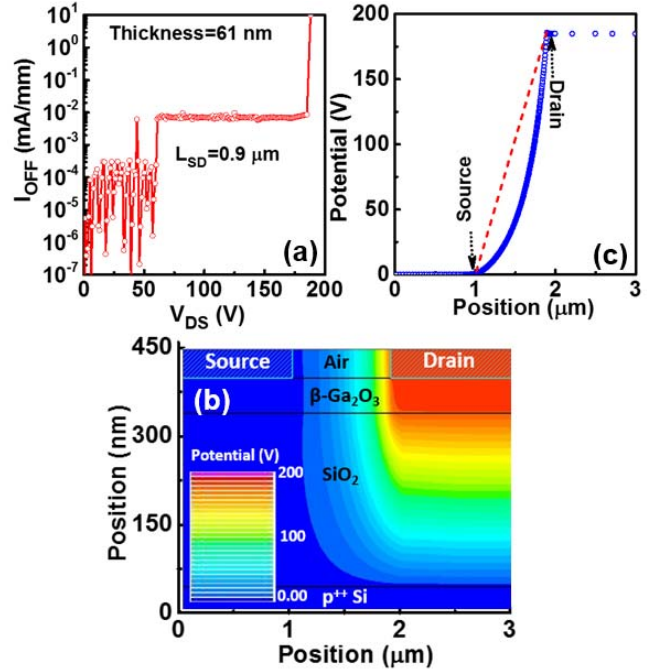

Fig. 5. (a) Off-state breakdown measurement of a floating gate GOOI FET with $L_{SD} = 0.9$ $\mu$m and $\beta$-Ga$_2$O$_3$ thickness = 61 nm. (b) Simulation of the electrostatic potential of the GOOI FET with $L_{SD} = 0.9$ $\mu$m and 300 nm SiO$_2$ gate dielectric. The same color is the equipotential contour. (c) The simulated potential along the source-$\beta$-Ga$_2$O$_3$-drain direction. Average $E_{av}$ = 2 MV/cm is obtained.

## IV. CONCLUSION

We have achieved record high $I_{DMAX}$ of 600/450 mA/mm for D/E-mode GOOI FETs. E-mode device can be realized through the thickness reduction of the $\beta$-Ga$_2$O$_3$ nano-membranes. High on/off ratio of $10^{10}$, low SS of 140 mV/dec and negligible $I_D$-$V_{GS}$ hysteresis reveals the high quality interface between $\beta$-Ga$_2$O$_3$ and SiO$_2$. E-mode GOOI FET with $L_{SD} = 0.9$ $\mu$m demonstrates a high BV = 185 V and $E_{av} = 2$ MV/cm, showing the great promise of GOOI FETs for future power devices.

## ACKNOWLEDGEMENT

The authors thank D. Jena for the valuable discussions and the technical support and guidance from the Sensors Directorate of Air Force Research Laboratory.